\documentclass[11pt,a4paper,notoc]{article}
\pdfoutput=1
\usepackage{jcappub}
\usepackage{slashed}
\DeclareMathOperator\erf{erf}
\title{The unbearable lightness of being: CDMS versus XENON}
\author[a]{Mads T.~Frandsen,}
\author[b]{Felix Kahlhoefer,}
\author[c]{Christopher McCabe,}
\author[b]{Subir Sarkar,}
\author[d]{and Kai Schmidt-Hoberg}
\affiliation[a]{CP$^3$-Origins and the Danish Institute for Advanced Study, University of Southern Denmark, Campusvej 55, DK-5230 Odense M, Denmark}
\affiliation[b]{Rudolf Peierls Centre for Theoretical Physics,
  University of Oxford, 1 Keble Road, Oxford OX1 3NP, United Kingdom}
\affiliation[c]{Institute for Particle Physics Phenomenology, Durham University, South Road, Durham, DH1 3LE, United Kingdom}
\affiliation[d]{Theory Division, CERN, 1211 Geneva 23, Switzerland}
\emailAdd{frandsen@cp3-origins.net}
\emailAdd{felix.kahlhoefer@physics.ox.ac.uk}
\emailAdd{christopher.mccabe@durham.ac.uk}
\emailAdd{s.sarkar@physics.ox.ac.uk}
\emailAdd{kai.schmidt-hoberg@cern.ch}
\arxivnumber{CP3-Origins-2013-012 DNRF90 , DIAS-2013-12, OUTP-13-09P, IPPP/13/23, DCPT/13/46, CERN-PH-TH/2013-081}
\abstract{The CDMS-II collaboration has reported 3 events in a Si detector, which are consistent with being nuclear recoils due to scattering of Galactic dark matter particles with a mass of $\sim 8.6$~GeV and a cross-section on neutrons of $\sim 2 \times 10^{-41}$~cm$^2$. While a previous result from the XENON10 experiment has supposedly ruled out such particles as dark matter, we find by reanalysing the XENON10 data that this is not the case. Some tension remains however with the upper limit placed by the XENON100 experiment, independently of astrophysical uncertainties concerning the Galactic dark matter distribution. We explore possible ways of ameliorating this tension by altering the properties of dark matter interactions. Nevertheless, even with standard couplings, light dark matter is consistent with both CDMS and XENON10/100.}
\date{\today}
\notoc
\begin{document}

\maketitle

\section{Introduction}

Identifying dark matter (DM) is  a key challenge for modern astro-particle physics~\cite{Sarkar:1990tf}. Its presence has been inferred solely through gravitational interactions but it is theoretically well-motivated to assume that DM consists of new particles arising in physics beyond the Standard Model (SM). It would then be natural for the strength of DM interactions to be less even than that of the weak interactions, consistent with the inferred collisionless behaviour of DM on cosmological scales. The relic abundance of weakly interacting massive particles (WIMPs) which decoupled from thermal equilibrium in the early universe is also naturally of order the observed DM abundance if they have a weak scale mass of ${\cal O}(100)$~GeV --- the so-called `WIMP miracle'. Another attractive possibility is that the DM particles inherited the same matter-antimatter asymmetry as baryons are observed to have --- their mass would then have to be of ${\cal O}(5)\,\text{GeV}$. In either case, the most direct way to identify DM is with well-shielded underground experiments which can detect  the energy deposited by a recoiling nucleus when a passing DM particle from the Galactic halo scatters on it.

Many such searches have been conducted for over three decades but the overwhelming majority have focussed on DM particles of mass $100 \lesssim m_{\chi}/\text{GeV} \lesssim 1000$, motivated simultaneously by the WIMP miracle and by the existence of a natural particle candidate viz.~the neutralino in supersymmetric extensions of the SM. Less attention has been paid to searching for lighter particles, both in the absence of as clear a theoretical motivation for these, as well as the increased difficulty in distinguishing nuclear recoils from background at low energies. However in recent years results from several experiments have  suggested $m_{\chi} \lesssim10$~GeV. This mass range is favoured by an excess of events over the expected background  observed by CoGeNT~\cite{Aalseth:2011wp} and CRESST-II~\cite{Angloher:2011uu} as well as the long-standing annual modulation signal seen by DAMA~\cite{Bernabei:2010mq}. However,  constraints set by other leading experiments, notably XENON10~\cite{Angle:2011th}, XENON100~\cite{Aprile:2012nq} and CDMS-Ge~\cite{Ahmed:2009zw, Ahmed:2010wy}, are in strong tension with these results and have cast doubt on the DM interpretation of these signals (see e.g.~\cite{Frandsen:2011ts, Farina:2011pw, Fox:2011px, McCabe:2011sr,Kopp:2011yr,Kelso:2011gd}).

The direct detection of `light' DM is particularly challenging, because it is hard to reliably determine the detector response to a potential DM signal at such low recoil energies~\cite{Savage:2010tg, Cline:2012ei}. Moreover, the interpretation of the data is  sensitive to the assumed properties (in particular, velocity distribution) of the DM halo~\cite{McCabe:2010zh,Green:2011bv,Fairbairn:2012zs,Strigari:2012gn}. To minimise these uncertainties, an ideal experiment to probe the low-mass region would have target nuclei with low mass number and a low energy threshold. Si detectors satisfy both criteria, making them an obvious choice for this purpose. With increasing interest amongst theorists in developing particle physics models for  light DM (and its co-genesis with baryons), the experimental community is becoming increasingly aware of the importance of looking in this mass region.

The CDMS-II collaboration has recently presented two analyses of data taken with Si detectors. While no candidate events were found in the small data set from the first `Five-Tower Run' of CDMS-II \cite{Agnese:2013cvt}, the longer exposure of 140.2 kg-days during 2007--08 \cite{Agnese:2013rvf} revealed 3 events in the DM search region, compared to a background expectation of $0.62$ events. Considering only the number of events the signal is compatible with background at the $\sim 2\sigma$ level, but taking into account their distribution, a background+DM interpretation of the data is preferred over the background-only hypothesis with a probability of 99.8\%. The highest likelihood is found for a DM particle with mass 8.6 GeV and cross-section $1.9 \times 10^{-41}$~cm$^2$, which is significantly below the DAMA and CRESST-II regions, but still in tension with bounds from XENON10/100.

We investigate under what conditions the tension between CDMS and XENON10/100 can be ameliorated or even resolved. In particular, we find the bound from the `S2-only analysis' of XENON10 to actually be \emph{weaker} by nearly a factor of 10 than the published value~\cite{Angle:2011th}. However, the tension with XENON100 \cite{Aprile:2012nq} remains and cannot easily be evaded by appealing to experimental or astrophysical uncertainties. We therefore consider various particle physics modifications of the interactions between DM and SM quarks. While a momentum- or velocity-dependence in the cross-section does not improve the situation, we find that \emph{isospin-dependent} couplings as well as \emph{exothermic} DM-scattering can bring XENON10/100 and CDMS into better agreement. We do not attempt, however, to find a common interpretation of the CDMS results and the signals found by DAMA, CoGeNT and CRESST-II.

In \S~\ref{directdetection}, we introduce the standard framework for interpreting DM signals and review both the CDMS best-fit region and the bounds from XENON10/100. We pay particular attention to the XENON10 bound and its dependence on the form of the ionisation yield $\mathcal{Q}_y$, which is \emph{essential} to understand the energy scale. In \S~\ref{vmin}, we briefly review $v_\text{min}$-space, which enables comparison of results from different direct detection experiments without having to assume a specific DM velocity distribution. We show that CDMS and XENON10/100 probe much the same region of $v_\text{min}$-space, hence the tension between them is \emph{independent} both of astrophysical uncertainties concerning the DM halo and any momentum- and velocity-dependence of the cross-section. We illustrate this by giving some examples for possible modifications. Then in \S~\ref{agreement} we consider possible changes in the underlying particle physics, which can suppress the sensitivity of Xe targets relative to Si, and thus bring CDMS and XENON10/100 into better agreement. We present our conclusions in \S~\ref{conclusions}.

\section{General framework and discussion of experimental results \label{directdetection}}

We first review the standard method of analysing DM direct detection experiments and then apply this to the silicon data from CDMS-II (CDMS-Si in short) and the XENON10/100 experiments. The event rate in a given detector depends on the flux of Galactic DM particles passing through the detector and the DM-nucleon scattering cross-section. In the laboratory frame the differential event rate with respect to recoil energy $E_\text{R}$ is given by
\begin{equation}
\frac{\text{d}R}{\text{d}E_\text{R}} = 
\frac{\rho}{m_{\text{N}} m_\chi}  \int_{v_\text{min}}^\infty 
v f(\boldsymbol{v} + \boldsymbol{v}_\text{E}(t)) \frac{\text{d} \sigma}{\mathrm{d} E_{\text{R}}} F^2(E_{\text{R}})\, \text{d}^3 v\; ,
\label{eq:dRdE}
\end{equation}
where $\rho$ is the local halo DM density, $m_\chi$ and $m_{\text{N}}$ are the DM and target nucleus mass respectively, $f(v)$ is the local DM velocity distribution evaluated in the Galactic rest frame, $v=|\boldsymbol{v}|$ and $\boldsymbol{v}_\text{E}(t)$ is the velocity of the Earth relative to the Galactic rest frame \cite{Gelmini:2000dm, Schoenrich:2009bx}. We have separated the differential scattering cross-section into the piece for scattering off a point like nucleus, $\mathrm{d} \sigma/\mathrm{d} E_{\text{R}}$, and the nuclear form factor $F(E_\text{R})$, which accounts for the finite size of the nucleus. For elastic scattering the \emph{minimum} velocity required for a DM particle to transfer the energy $E_\text{R}$ to a given nucleus $N$ is:
\begin{equation}\label{eq:vmin}
v_\text{min}(E_\text{R}) = \sqrt{\frac{m_\text{N} E_\text{R}}{2 \mu^2}} \;,
\end{equation}
where $\mu$ is the reduced mass of the DM-nucleus system.

For a detector with energy resolution $\Delta E_\text{R}$, the number of events $N(E_1, E_2)$ expected in an energy range $\left[E_1, E_2\right]$ is:
\begin{equation}
N(E_1, E_2) = \text{Ex} \, \int \text{Res} (E_1, E_2, E_\text{R}) \, \epsilon(E_\text{R}) \, \frac{\text{d}R}{\text{d}E_\text{R}} \, \text{d}E_\text{R} \;,
\end{equation}
where $\text{Ex}$ is the exposure of the detector, $\epsilon(E_\text{R})$ is the detector acceptance and
\begin{equation}
\text{Res} (E_1, E_2, E_\text{R}) = \frac{1}{2}\left[\erf\left(\frac{E_2 - E_\text{R}}{\sqrt{2} \Delta E_\text{R}}\right)-\erf\left(\frac{E_1 - E_\text{R}}{\sqrt{2} \Delta E_\text{R}}\right)\right]
\end{equation}
is the detector response function~\cite{Savage:2008er}.

\begin{figure}[t]
\centering
\includegraphics[width=0.6 \textwidth]{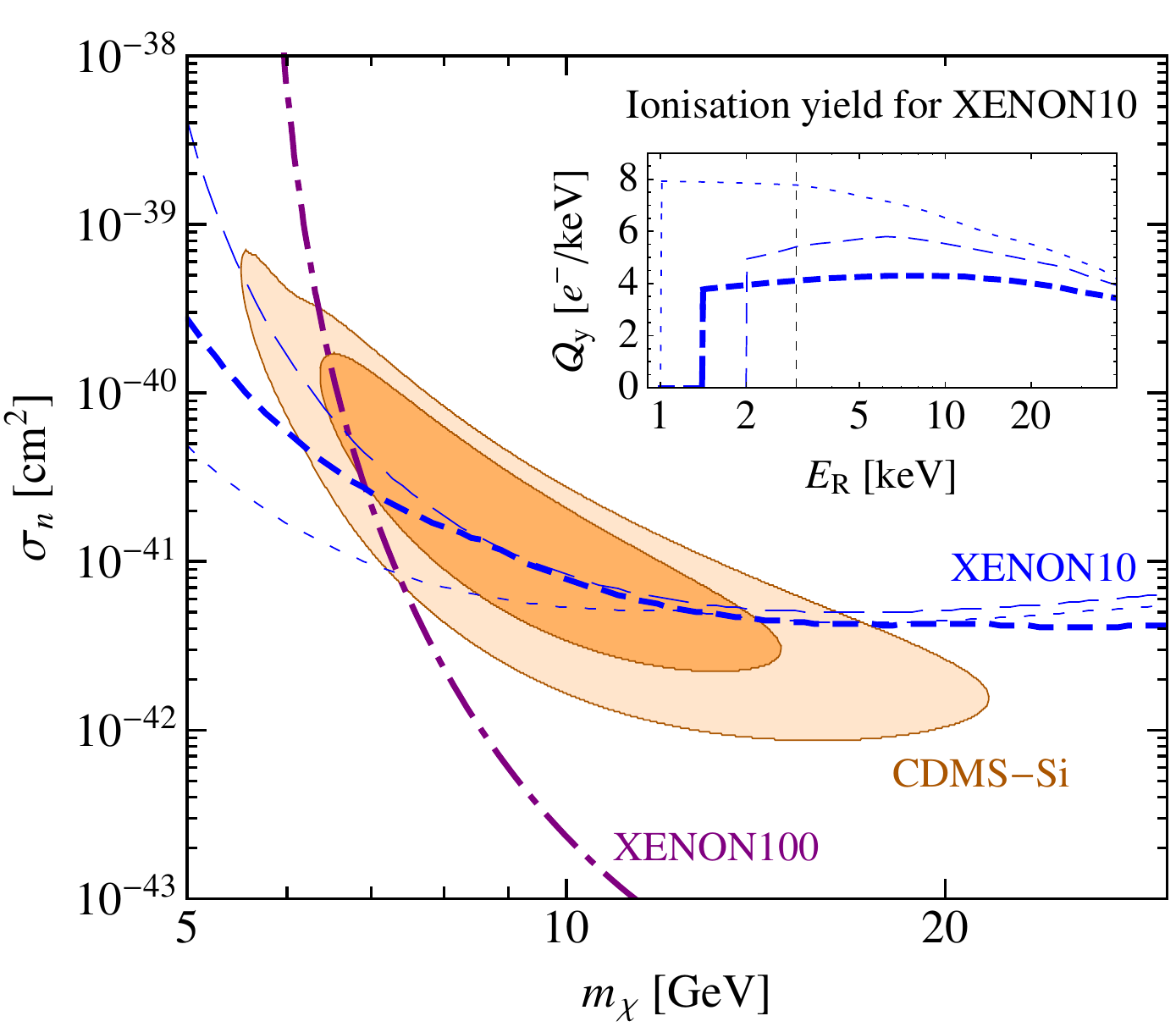}
\caption{CDMS-Si confidence region (68\% and 90\% CL) together with the 90\% exclusion curves from XENON10 and XENON100 from our analysis, assuming standard elastic spin-independent scattering and equal coupling to protons and neutrons. Our CDMS-Si confidence region and XENON100 bound agree well with the results from the respective collaborations \cite{Agnese:2013rvf, Aprile:2012nq}, however, our XENON10 bound is significantly weaker than the published one~\cite{Angle:2011th}. We consider three choices of the ionisation yield $\mathcal{Q}_\text{y}$ at low energy to illustrate the corresponding variation  of the extracted bound.}
\label{fig:standard}
\end{figure}

We now focus on whether CDMS-Si is compatible with XENON10 and XENON100. We consider only spin-independent scattering in this work since only a very small fraction of naturally occurring Si contains an isotope with nuclear spin. The usual parameterisation for a spin-independent cross-section is
\begin{equation}
\label{eq:usualSI}
\frac{\text{d}\sigma}{\text{d} E_\text{R}} = C^2_\text{T} (A,Z)  \frac{m_\text{N} \sigma_n}{2 \mu_{n\chi}^2 v^2}\;,
\end{equation}
where $\mu_{n\chi}$ is the reduced DM-nucleon mass, $C_\text{T}(A, Z) \equiv \left(f_p/f_n Z + (A-Z)\right)$, $A$ and $Z$ are the mass and charge number of the target nucleus and $f_{n,p}$ denote the effective DM coupling to neutrons and protons, respectively. Experiments typically quote results in terms of $\sigma_n$, the DM-neutron cross-section at zero momentum transfer. The above parameterisation of the cross-section holds e.g.\ for DM-nucleus scattering mediated by a heavy $CP$-even scalar boson or by a heavy vector mediator. However, other interactions can give rise to a cross-section that \emph{cannot} be parameterised in this way. We consider two such examples in \S~\ref{vmin}.

Unless otherwise stated, we adopt for the astrophysical parameters the Standard Halo Model (SHM), i.e.~a Maxwell-Boltzmann distribution for $f(v)$ with $v_0=220$~km\,s$^{-1}$ and $v_{\rm{esc}}=544$~km\,s$^{-1}$, and take $\rho=0.3~\text{GeV\,cm}^{-3}$. We use the Helm form factor \cite{Lewin:1995rx} and take $f_n/f_p=1$. The results of our calculations for this choice of parameters are shown in Fig.~\ref{fig:standard}. 

For XENON100 we take the data from an exposure of $20.9\,\text{kg-years}$~\cite{Aprile:2012nq}, combined with the recent measurements of $\mathcal{L}_\text{eff}$ \cite{Aprile:2013teh}. We calculate the energy resolution under the assumption that it is dominated by Poisson fluctuations in the number of photoelectrons and obtain exclusion limits using the `maximum gap method'~\cite{Yellin}. As noted already, CDMS-Si found 3 events after background rejection in data from an exposure of $140.2\,\text{kg-days}$~\cite{Agnese:2013rvf}. For our analysis we assume an energy resolution of $\Delta E_\text{R} = 0.3\,\text{keV}$~\cite{Akerib:2010pv} and use the detector acceptance from~\cite{Agnese:2013rvf}. We take the normalised background distributions from~\cite{McCarthy} and rescale the individual contributions in such a way that 0.41, 0.13 and 0.08 events are expected from surface events, neutrons and $^{206}$Pb, respectively. 

To calculate the allowed parameter region, we employ the extended maximum likelihood method~\cite{Barlow1990496}. The contours we show correspond to $68\%$ and $90\%$ confidence level (CL) --- these agree well with the ones shown by the CDMS-II collaboration~\cite{Agnese:2013rvf}. In excellent agreement with their analysis, we find that the best-fit point is $m_\chi = 8.55\,\text{GeV}$ with $\sigma_n = 2\times10^{-41}\,\text{cm}^2$, and that the background+DM hypothesis is favoured over the background-only hypothesis with a probability of $99.6\%$. We find that varying the normalisation of the background within the range suggested in~\cite{Agnese:2013rvf} while keeping its shape fixed does not sensibly affect either the significance of the background+DM hypothesis, nor the shape of the best-fit region.

For XENON10, we use the S2-only analysis~\cite{Angle:2011th}. The sensitivity of this analysis depends strongly on the behaviour of the ionisation yield $\mathcal{Q}_\text{y}$ at low energies~\cite{Cline:2012ei}. To calculate the central bound (thick dashed blue line) in Fig.~\ref{fig:standard} we adopt the original choice of $\mathcal{Q}_\text{y}$ made in~\cite{Angle:2011th}, assuming that $\mathcal{Q}_\text{y}$ vanishes for $E_\text{R} < 1.4\,\text{keV}$. We assume that the energy resolution is given by $\Delta E_\text{R} = E_\text{R} / \sqrt{E_\text{R} \mathcal{Q}_\text{y}(E_\text{R})}$ and use the maximum gap method to set a bound. We have checked that the `$p_\text{max}$ method'~\cite{Yellin} and the `binned Poisson method'~\cite{Green:2001xy, Savage:2008er} give comparable bounds. In every case, we find bounds that are \emph{significantly weaker} than the published one~\cite{Angle:2011th} --- even when we assume exactly the same values of $\mathcal{Q}_\text{y}$ --- but in good agreement with other independent analyses~\cite{Farina:2011pw,Cline:2012ei, Schwetz,Winkler}.\footnote{This conclusion is not affected by the different choices of astrophysical parameters, which we discuss in more detail in the following section.} It has been suggested that the XENON10 data be analysed by converting recoil spectra to photoelectrons and using Poisson statistics to model the detector response~\cite{Gondolo:2012rs}. We find that this procedure gives even weaker bounds at low DM mass. Therefore we assume Gaussian fluctuations in order to enable fair comparison with other results in the literature.

We also consider values of $\mathcal{Q}_\text{y}$ that have recently been determined by the XENON100 collaboration by comparing calibration data to Monte Carlo simulations. We adopt the range of values given in Fig.~3 of~\cite{Aprile:2013teh}, which we have reproduced together with the choice of $\mathcal{Q}_\text{y}$ from~\cite{Angle:2011th} in Fig.~\ref{fig:standard} (inset). Most of the sensitivity for detecting low-mass DM relies on recoil energies below 3~keV, where $\mathcal{Q}_\text{y}$ cannot be reliably extracted from the data. We therefore extrapolate $\mathcal{Q}_\text{y}$ in two different ways that bracket the range of possibilities: the pessimistic choice (light dashed blue line in Fig.~\ref{fig:standard}) assumes that $\mathcal{Q}_\text{y}$ follows the lower bound of the range given in~\cite{Aprile:2013teh} and then drops to zero at 2~keV. The optimistic choice (light dotted blue line in Fig.~\ref{fig:standard}) assumes that $\mathcal{Q}_\text{y}$ follows the upper bound of the range, continuing to rise up to $8\,e^-/\text{keV}$ below $3\,\text{keV}$, and then drops to 0 at 1~keV. Even for the optimistic case we do not extrapolate $\mathcal{Q}_\text{y}$ below 1~keV, i.e.~we follow the approach of~\cite{Angle:2011th} and neglect upward fluctuations. Changing $\mathcal{Q}_\text{y}$ also changes the energy scale shown in Fig.~2 of~\cite{Angle:2011th} so we convert the energies of the observed events appropriately when setting the bound. It is clear from Fig.~\ref{fig:standard} that although the XENON10 S2-only analysis does constrain the CDMS-Si parameter region, an unambiguous bound cannot be placed at low mass because of uncertainty in the value of the ionisation yield $\mathcal{Q}_\text{y}$ at low energy. 

Henceforth we present only one XENON10 bound, adopting the original choice of $\mathcal{Q}_\text{y}$ from~\cite{Angle:2011th}. As this is significantly weaker than the published bound~\cite{Angle:2011th}, CDMS-Si and XENON10/100 are actually \emph{consistent} at about $90\%$ CL. Nevertheless there is some tension between these results. We will now explore possible ways to ameliorate this tension.

\section[Analysing the experiments in vmin-space]{Analysing the experiments in $v_\text{min}$-space}\label{vmin}

We have seen that many parameters need to be specified before a theoretical prediction for the number of scattering events in a direct detection experiment can be compared with the observed number. While a parameter such as the local DM density affects all experiments in the same way, other parameters can change the number of events in one experiment while having no impact on another experiment. A useful technique to gain insight into this involves mapping the experimental result into $v_\text{min}$-space~\cite{Fox:2010bz}. If experiments probe different regions of this space, they will be affected \emph{differently} by varying parameters such as the local escape velocity $v_\text{esc}$; conversely, if experiments probe the same region of $v_\text{min}$-space, then modifying such parameters cannot improve agreement between the experiments. We first apply this technique in the usual way to astrophysical parameters, before applying it also to momentum-dependent interactions. Our discussion and notation closely follow~\cite{Frandsen:2011gi}.

\subsection{Varying astrophysical parameters}

\begin{figure}[t]
\centering
\includegraphics[width=0.49\textwidth]{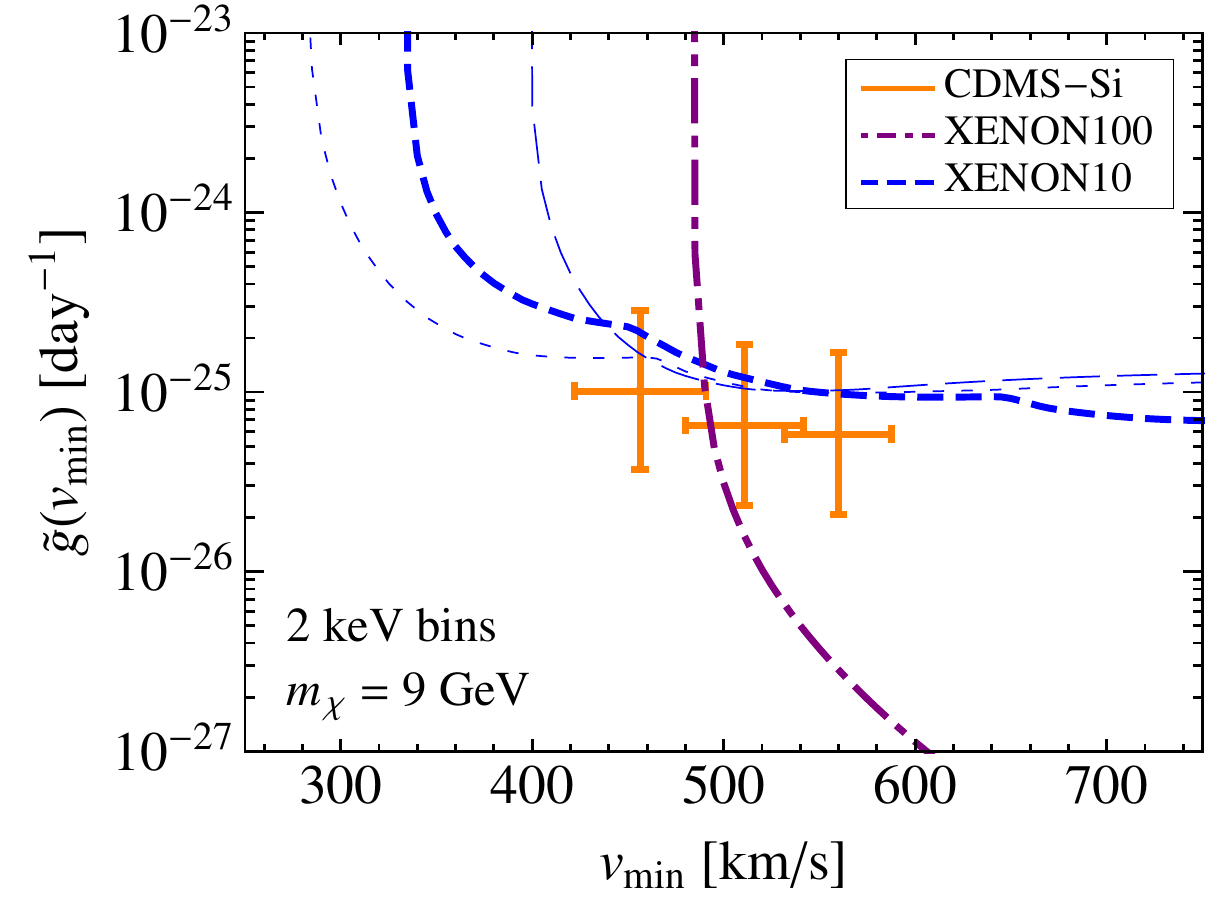}
\includegraphics[width=0.49\textwidth]{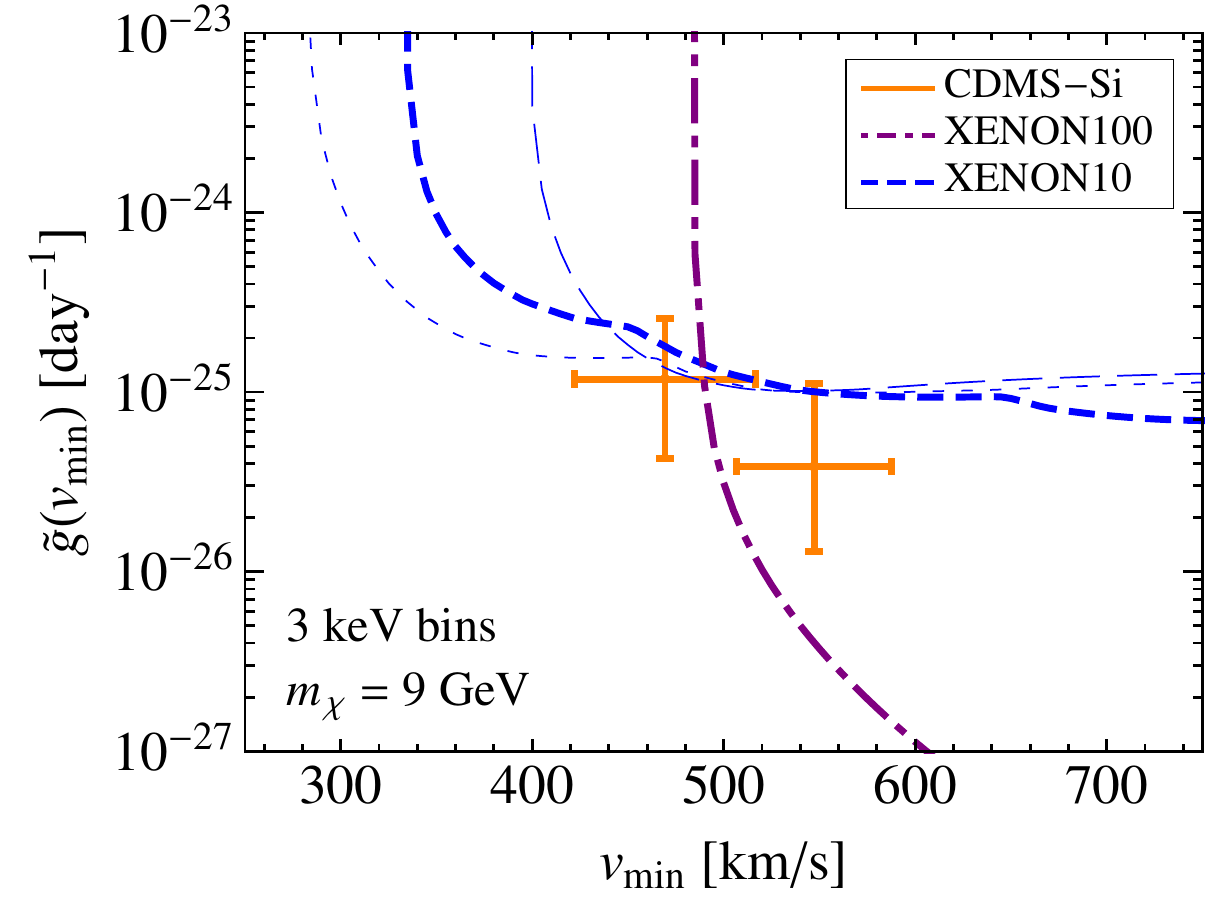}
\includegraphics[width=0.49\textwidth]{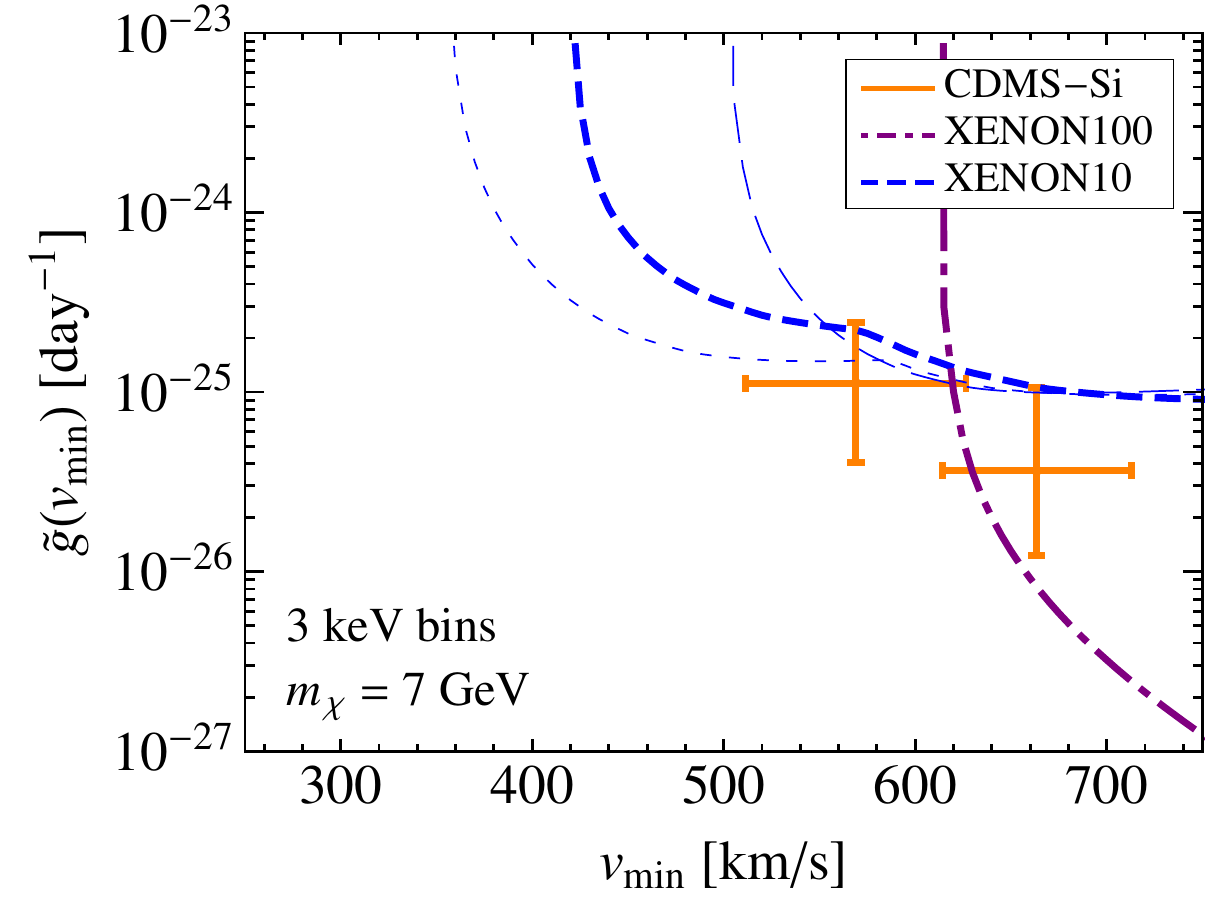}
\includegraphics[width=0.49\textwidth]{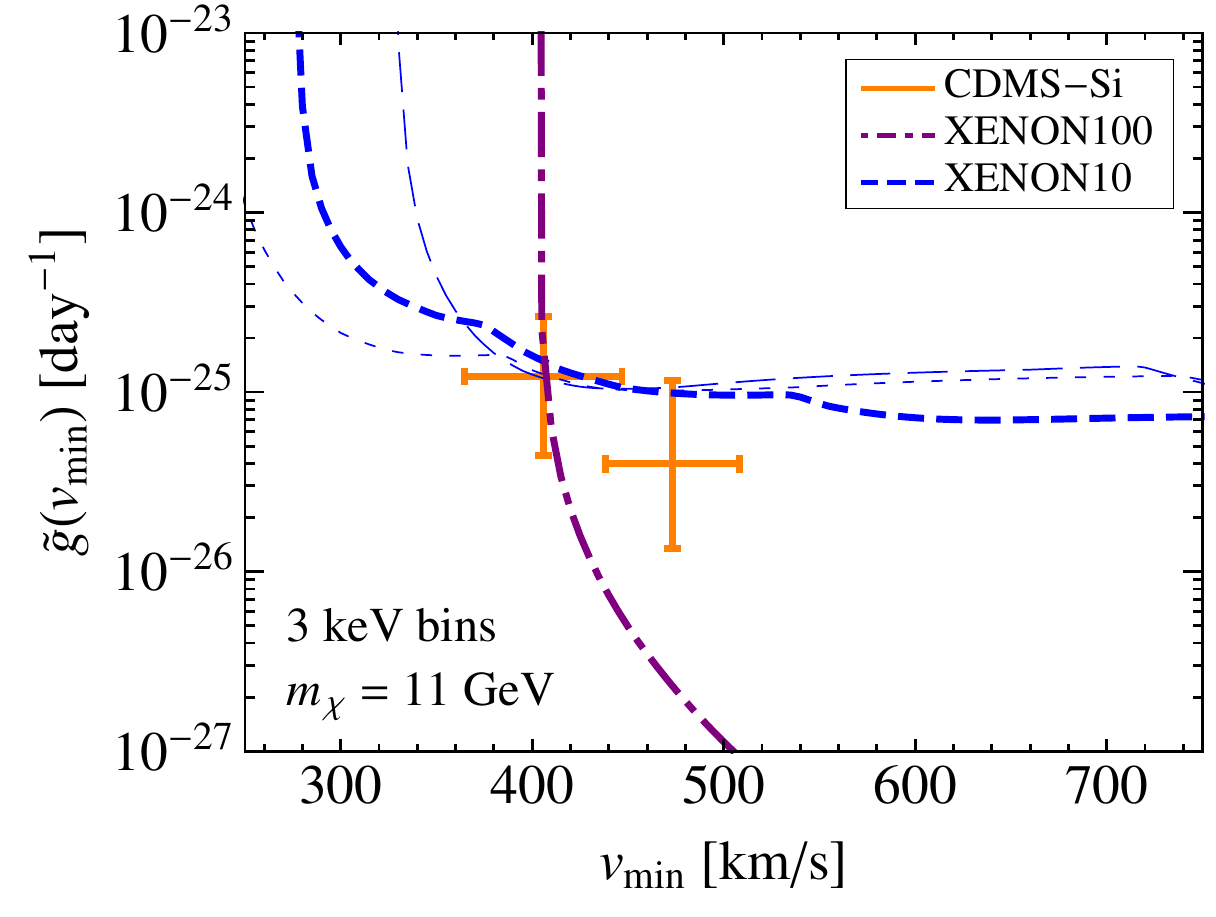}
\caption{The CDMS-Si and XENON10/100 results translated into $v_\text{min}$-space. The upper panels show the case $m_\chi=9$~GeV for two choices of binning. In the left (right) panel the bin width is $2\,\text{keV}$ ($3\,\text{keV}$). The choice of binning does not alter our conclusions. For all the cases considered, the region of $v_\text{min}$-space probed by CDMS-Si is constrained by XENON10/100.}
\label{fig:vmin}
\end{figure}

After substituting the usual parameterisation of the cross-section for spin-independent scattering from Eq.~\eqref{eq:usualSI} into Eq.~\eqref{eq:dRdE}, we see that direct detection experiments do not directly probe the local velocity distribution $f(v)$, but rather the velocity integral $g(v_\text{min})=\int_{v_\text{min}} f(v)/v \,\text{d}^3v$, where $v_\text{min}$ is defined by Eq.~\eqref{eq:vmin}. For our purposes, it will be convenient to absorb the DM mass, cross-section and density into this definition and consider the `rescaled velocity integral':
\begin{equation}
\tilde{g}(v_\text{min}) = \frac{\rho \, \sigma_n}{m_\chi} \, g(v_\text{min}) \; .
\label{eq:dRdEnorm}
\end{equation}
Since $g(v_\text{min})$ is a monotonically decreasing function of $v_\text{min}$~\cite{Fox:2010bu}, we can use experimental null results to place upper bounds on $g(v_\text{min})$ for $v_\text{min} = \hat{v}_\text{min}$ using the standard techniques for setting an exclusion limit, but with~\cite{Fox:2010bz, Frandsen:2011gi}
\begin{equation}
\tilde{g}(v_\text{min})= \tilde{g}(\hat{v}_{\rm{min}})\Theta(\hat{v}_{\rm{min}}-v_\text{min})\;.
\end{equation}
The resulting bounds from XENON10 and XENON100 are shown as the blue (dashed) and purple (dot-dashed) lines in Fig.~\ref{fig:vmin}, respectively. 

In order to deduce information on $\tilde{g}(v_\text{min})$ from CDMS-Si, we need to infer the differential event rate and use the formula:
\begin{equation}
\tilde{g}(v_\text{min})=\frac{2 \mu_{n \chi}^2}{C_T^2(A,Z)F^2(E_{\rm{R}})}\frac{\mathrm{d} R}{\mathrm{d} E_{\text{R}}}\;.
\end{equation}
The differential event rate is found most easily by binning both the observed events and the background expectation. We subtract the background using the Feldman-Cousins technique~\cite{Feldman:1997qc}, which allows us to calculate an upper and lower bound on the signal event rate at $1\sigma$ CL. The orange data points in Fig.~\ref{fig:vmin} show the result of this procedure for $m_{\chi}=9\,\text{GeV}$ (upper panels) and $m_{\chi}=7\,\text{GeV}$ and $m_{\chi}=11\,\text{GeV}$ (lower left and lower right panels respectively). Binning the data introduces a certain arbitrariness so we check the robustness of our results by considering two choices of the bin width: $2\,\text{keV}$ and $3\,\text{keV}$ for the upper left and right panels of Fig.~\ref{fig:vmin} respectively. The inferred values for $\tilde{g}(v_\text{min})$ agree well, implying that our conclusions are largely independent of the choice of bin width. In all cases, the highest bin is in significant tension with the XENON100 bound except for the case $m_{\chi}=7\,\text{GeV}$, corresponding to the least constrained mass in Fig.~\ref{fig:standard}.

We observe from Fig.~\ref{fig:vmin} that all three experiments probe essentially the same region of $v_\text{min}$-space. This suggests that it will \emph{not} be possible to significantly improve the consistency of CDMS-Si and XENON10/100 by varying astrophysical parameters. To explicitly demonstrate that this is so, we consider two variations in astrophysical parameters. 
\begin{figure}[t]
\centering
\includegraphics[width=0.49\textwidth]{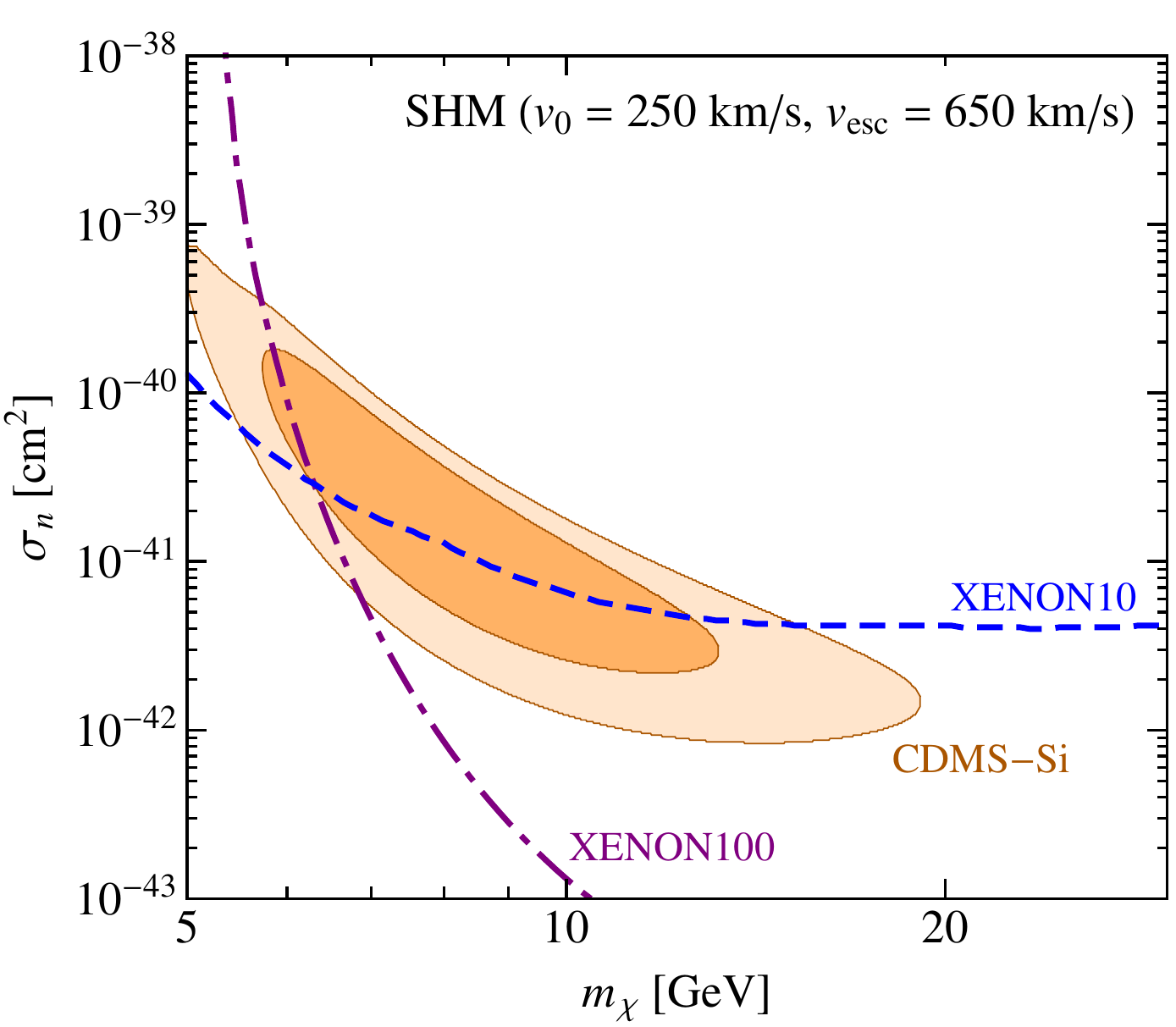}
\includegraphics[width=0.49\textwidth]{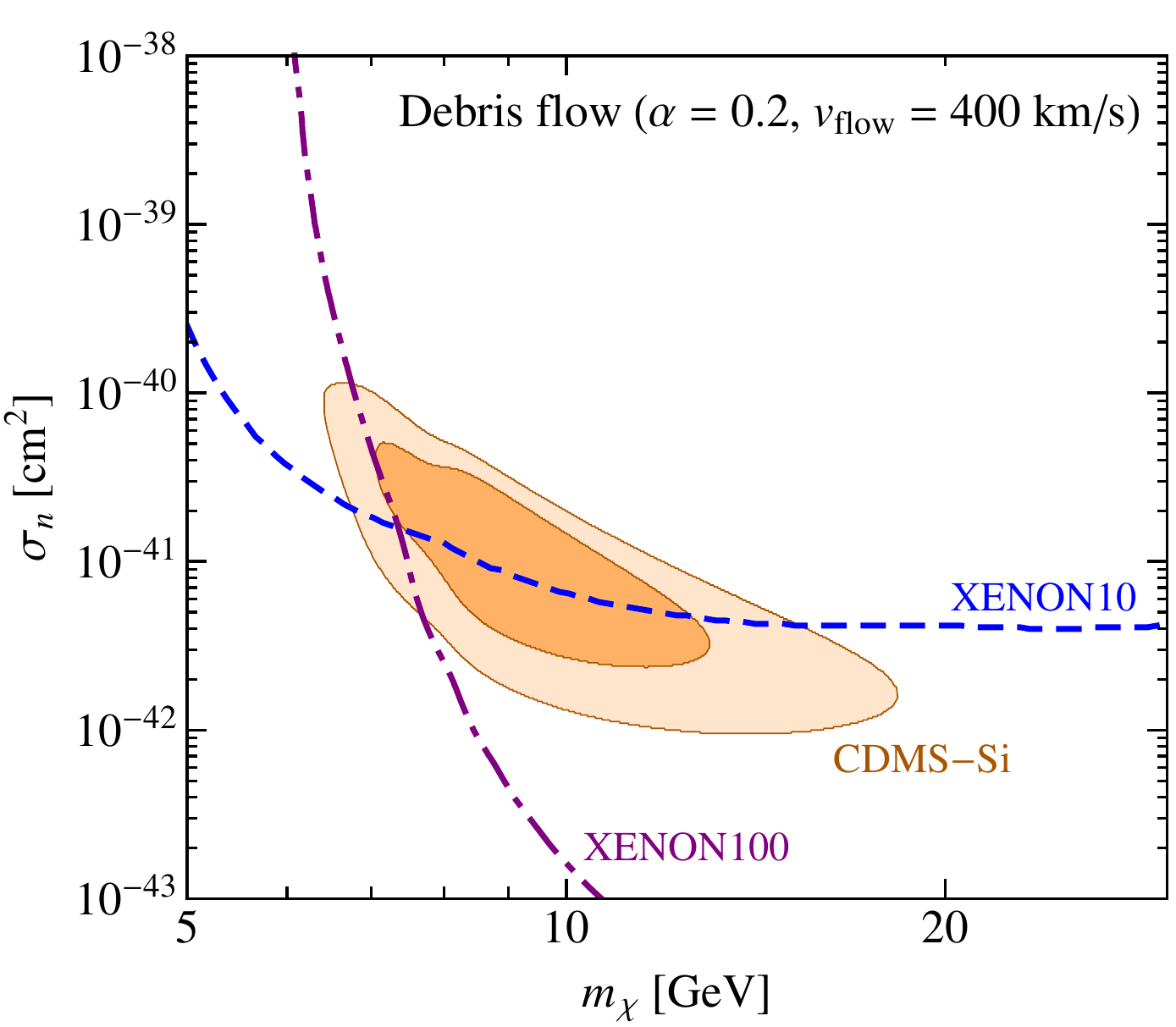}
\caption{The CDMS-Si region and XENON10/100 bounds calculated for different astrophysical parameters. The left panel shows the SHM, but with higher values of $v_0$ and $v_\text{esc}$. The right panel shows the effect of a debris flow which contributes 20\% to the local DM density. Although the signal region and bounds move by a few GeV, the consistency of CDMS-Si and XENON10/100 is unchanged in either case, since all experiments probe the same region of $v_\text{min}$ space.}
\label{fig:changeast}
\end{figure}
In the left panel of Fig.~\ref{fig:changeast} we keep the usual Maxwell-Boltzmann velocity distribution but choose $v_0=250$~km/s and $v_{\rm{esc}}=650$~km/s, which are at the upper end of the allowed range for these parameters (see e.g.~\cite{McCabe:2010zh} and references within). Although we see that the CDMS-Si region and XENON10/100 bounds move towards lower masses by $\sim1\,\text{GeV}$, the tension between the experiments remains essentially unchanged. As a more radical modification we consider the effect of a `debris flow' in the presence of which $g(v_\text{min})$ is given by~\cite{Lisanti:2011as,Kuhlen:2012fz}
\begin{equation}
g(v_\text{min}) = (1 - \alpha) \, g(v_\text{min})_\text{halo} + \alpha \times \left\{
     \begin{array}{ll}
       \frac{1}{v_\text{flow}} & \text{if } v_\text{min} < (v_\text{flow} - v_\text{E})\\
       \frac{v_\text{flow} + v_\text{E} - v_\text{min}}{2 \, v_\text{flow} \, v_\text{E}} & \text{if } (v_\text{flow} - v_\text{E}) < v_\text{min} < (v_\text{flow} + v_\text{E})\\
       0 & \text{if } v_\text{min} > (v_\text{flow} + v_\text{E}) \end{array}
   \right.\;,
\end{equation}
where $v_\text{flow}$ is the velocity of the flow in the Galactic rest frame and $\alpha$ is the contribution of the debris to the local DM density. In the right panel of Fig.~\ref{fig:changeast} we have chosen $\alpha = 0.2$ and $v_\text{flow} = 400\,\text{km/s}$. We  see again that although the preferred mass changes, the overall agreement does not.

\subsection{Non-standard momentum and velocity dependence}

We have seen that CDMS-Si and XENON10/100 probe the same region of $v_\text{min}$-space so the tension between these experiments is independent of astrophysical uncertainties. We now consider the effect of modifying the particle physics to introduce an additional momentum- or velocity-dependence of the cross-section which can change the recoil energy spectrum. 

Non-trivial momentum dependence can be studied using the simple parameterisation~\cite{Chang:2009yt}
\begin{equation}
\frac{\mathrm{d}\sigma}{\mathrm{d}E_\mathrm{R}} = \left(\frac{\mathrm{d}\sigma}{\mathrm{d}E_\mathrm{R}}\right)_0 \left(\frac{q^2}{q^2_\text{ref}}\right)^n
\end{equation}
where $(\mathrm{d}\sigma/\mathrm{d}E_\mathrm{R})_0$ is the differential cross-section introduced in Eq.~\eqref{eq:usualSI}, $q = \sqrt{2 \, m_\mathrm{N} \, E_\mathrm{R}}$ is the momentum transfer and $q_\text{ref}$ is an arbitrary normalisation. 
\begin{figure}[!t]
\begin{center}
\includegraphics[width=0.45 \textwidth]{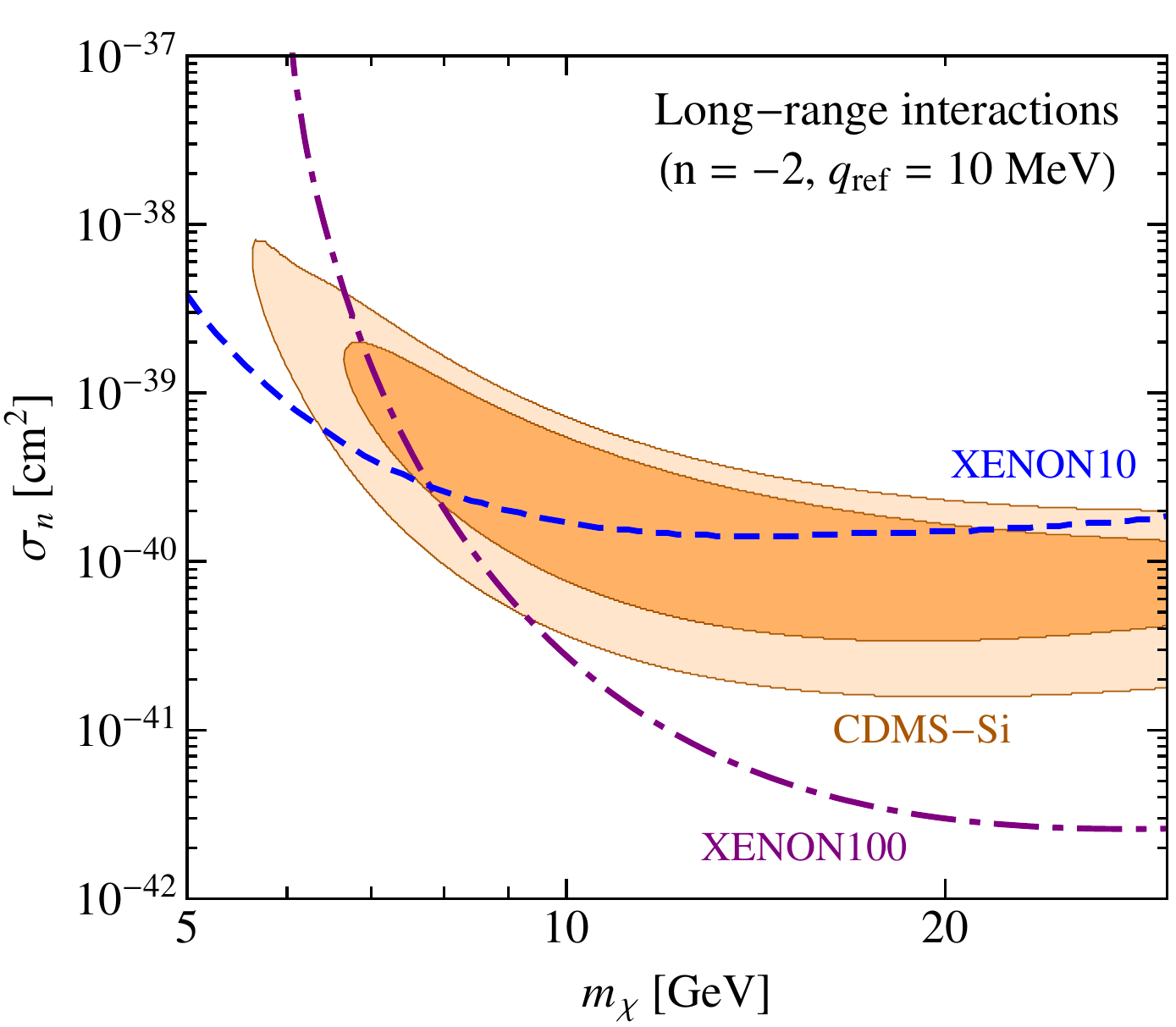}
\quad
\includegraphics[width=0.45 \textwidth]{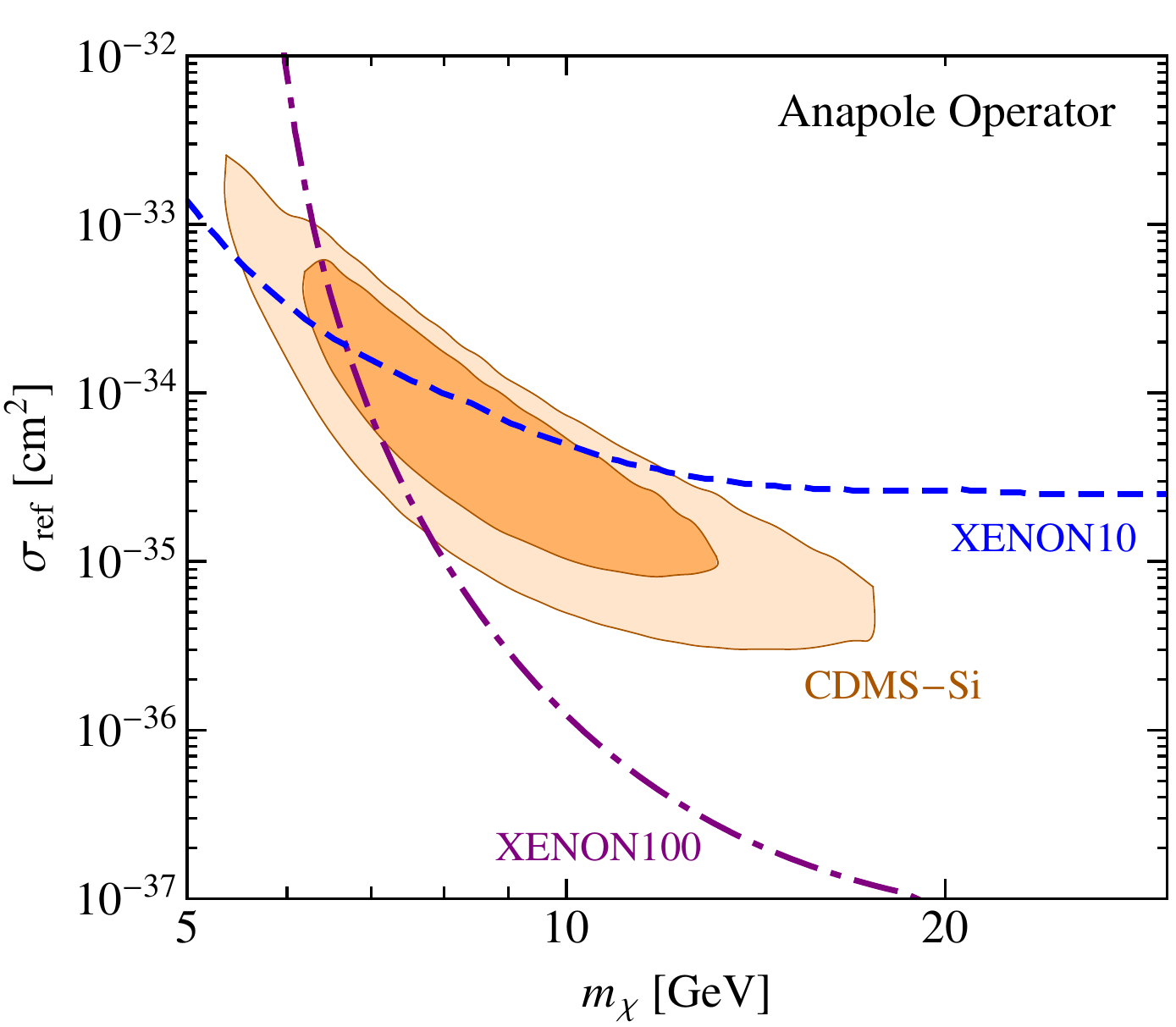}
\vspace{1mm}
\caption{\label{fig:momentum}Two examples for modifications of the momentum- and velocity-dependence of the differential cross-section. Left: Long-range interactions, which enhance the cross-section for small momentum transfer. Right: Anapole interactions, which suppress the cross-section for small momentum transfer and small velocity (note the change of vertical scale).}
\end{center}
\end{figure}
The crucial observation is that the momentum transfer $q$ is related to the minimal velocity $v_\text{min}$ by
\begin{equation}
v_\text{min} = \frac{q}{2 \mu} \; .
\end{equation}
For light DM, $m_\chi \ll m_\mathrm{N}$, hence $\mu \approx m_\chi$ independent of $m_\mathrm{N}$. Consequently, experiments that probe the same range of $v_\text{min}$ will also probe the same range of $q$ (see also~\cite{Feldstein:2009tr}). We  cannot therefore significantly shift the parameter region favoured by CDMS-Si relative to the XENON10/100 bounds by introducing a momentum dependence. This is illustrated in the left panel of Fig.~\ref{fig:momentum} for $q_\text{ref} = 10\,\text{MeV}$ and $n = -2$, which corresponds to long-range interactions arising from a (nearly) massless mediator. As expected, no significant improvement is found.

In many realistic models, the cross-section depends not only on the momentum transfer but also on the velocity $v$ of the incoming DM particle. This is the case e.g.\ for dipole interactions~\cite{Barger:2010gv,Banks:2010eh,DelNobile:2012tx} and anapole interactions~\cite{Fitzpatrick:2010br, Ho:2012bg}. Since we know already that both CDMS-Si and XENON10/100 probe the same part of the DM velocity distribution, such an additional velocity dependence is not expected to change the picture significantly. We confirm this expectation for the case of anapole interactions, which lead to a differential scattering cross-section~\cite{Ho:2012bg}:
\begin{equation} 
\frac{\mathrm{d}\sigma}{\mathrm{d} E_\mathrm{R}} = \frac{m_\mathrm{N} \, \sigma_\text{ref}}{2 \, \mu^2} \frac{Z^2}{v^2} \left[v^2 + \frac{q^2}{2 m_\mathrm{N}^2} \left(1 - \frac{m_\mathrm{N}^2}{2 \mu^2}\right)\right] \; .
\end{equation}
Our results are shown in the right panel of Fig.~\ref{fig:momentum}. As expected, because of the momentum- and velocity-suppression, the CDMS-Si favoured parameter region and the XENON10/100 bounds are moved to much larger cross-sections, but their relative position remains unchanged. For the anapole operator we do in fact observe a slight shift of the CDMS-Si region compared to the XENON10/100 bounds. This shift can be traced back to the fact that for anapole interactions DM particles couple to protons only leading to a factor of $Z^2$ rather than $A^2$ in the cross-section. We explore the effect of different DM couplings to protons and neutrons in more detail in the next section.

\section{Reducing the tension between CDMS-Si and XENON10/100}\label{agreement}

As we have seen CDMS-Si and XENON10/100 cannot be brought into better agreement by modifying either the DM velocity distribution or  the velocity/momentum dependence of the cross-section. To weaken the constraints from XENON10/100, we need to reduce the enhancement of the cross-section for heavy nuclei. In this section, we discuss two possible modifications of DM interactions that can increase the sensitivity of light targets compared to heavy ones: inelastic DM~\cite{TuckerSmith:2001hy} and isospin-dependent couplings~\cite{Kurylov:2003ra,Giuliani:2005my,Chang:2010yk,Feng:2011vu}.

\begin{figure}[!t]
\begin{center}
\includegraphics[width=0.45 \textwidth]{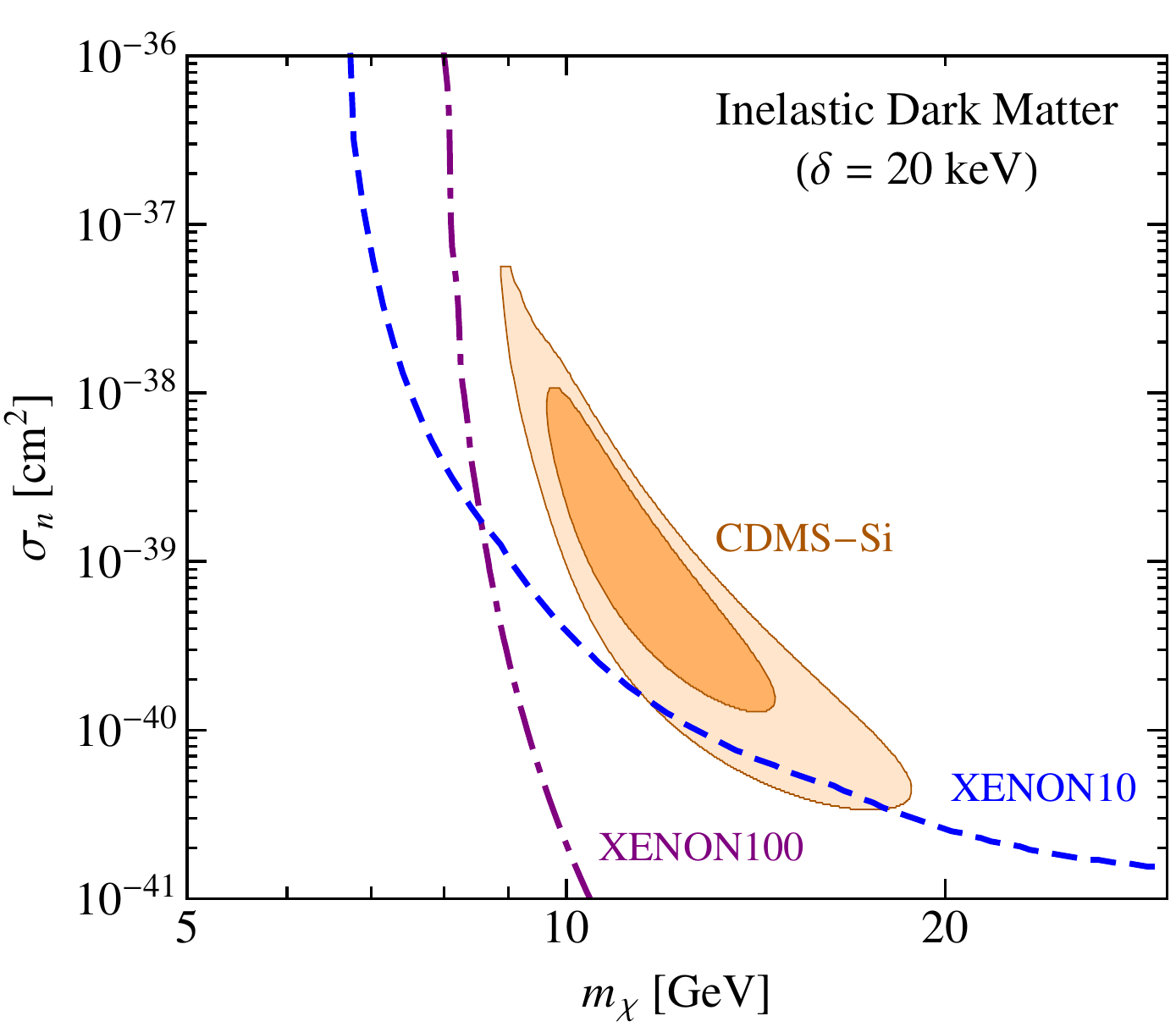}
\quad
\includegraphics[width=0.45 \textwidth]{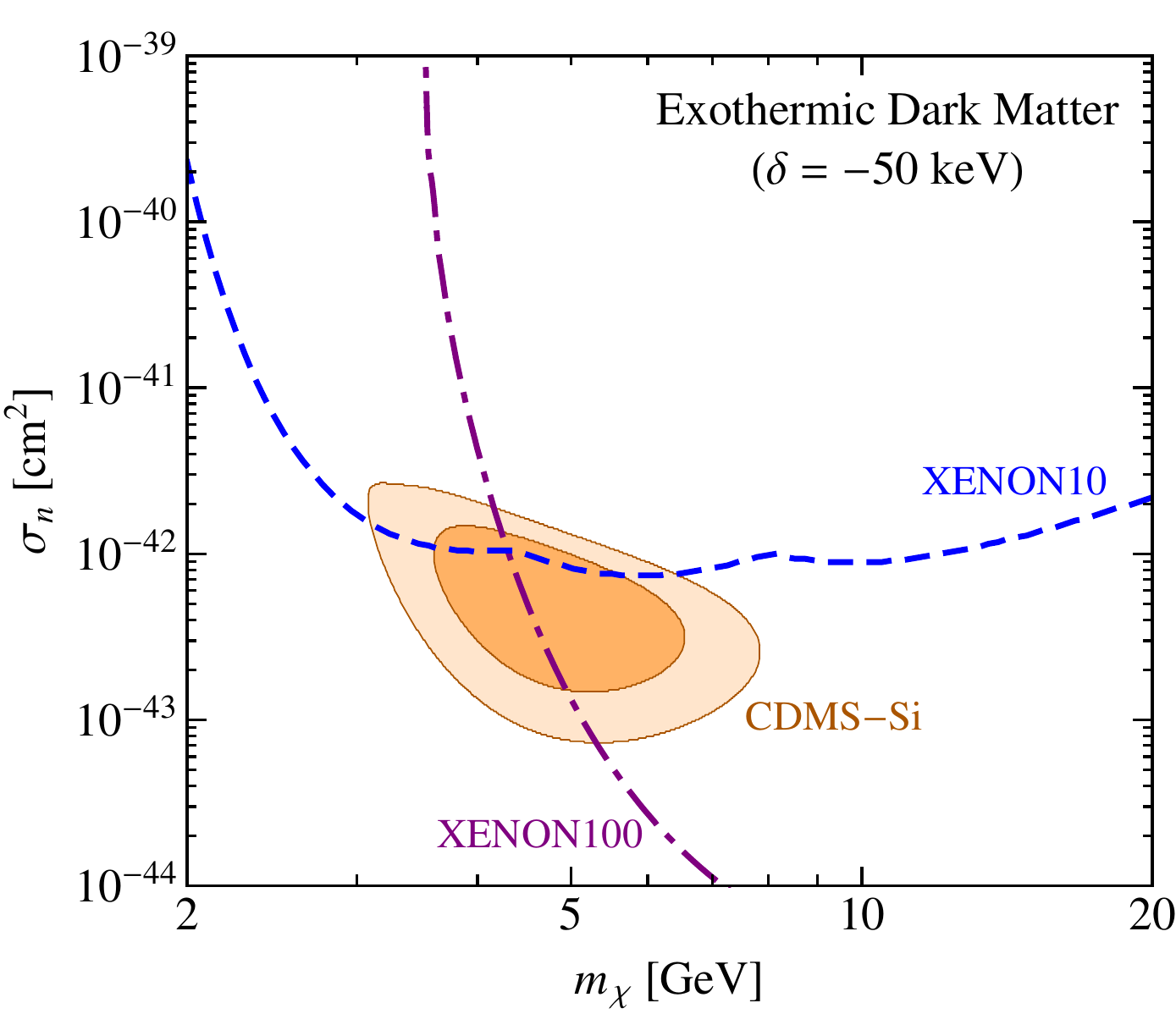}
\vspace{1mm}
\caption{\label{fig:inelastic} Two examples for inelastic scattering. Left: Inelastic interactions with $\delta > 0$, which suppress the scattering rate on light target nuclei. Right: Exothermic interactions with $\delta < 0$, which enhance the scattering rate on light target nuclei. Note the change of scales in these figures.}
\end{center}
\end{figure}

In the former case, DM-nucleon interactions require the transition between two DM states of slightly different mass. The minimum velocity required for a recoil of energy $E_\mathrm{R}$ is then:
\begin{equation}
v_\text{min} = \left|\delta + \frac{m_\mathrm{N} \, E_\mathrm{R}}{\mu} \right| \frac{1}{\sqrt{2 \, E_\mathrm{R} \, m_\mathrm{N}}} \; ,
\end{equation}
where $\delta$ is the mass splitting between incoming and outgoing DM state. If the incoming DM particle is lighter ($\delta > 0$), scattering will be enhanced for heavy targets. However, if only the heavier state is populated initially ($\delta < 0$), lighter targets will be favoured. This second case, referred to as exothermic DM~\cite{Graham:2010ca, Essig:2010ye}, thus seems a promising option to fully reconcile CDMS-Si and XENON10/100. We present our results in Fig.~\ref{fig:inelastic}. As expected, the XENON10/100 bounds are strengthened for inelastic DM and weakened for exothermic DM. In fact, a relatively small splitting of $\delta = -50\,\text{keV}$ is sufficient to bring CDMS-Si and XENON10/100 into good agreement.

Another possible route to concordance is based on the observation that the strength of the XENON10/100 bounds results partially from the factor $A^2$ in the cross-section, following from the assumption that DM couples equally to protons and neutrons. If DM couples to protons only (as for anapole interactions), the cross-section will be proportional to $Z^2$, thus favouring targets with a larger ratio of protons to neutrons. If the coupling to neutrons is slightly negative, $f_n / f_p < 0$, targets with a large fraction of neutrons will suffer more strongly from destructive interference between protons and neutrons. Such negative values of $f_n / f_p$ can arise e.g.\ in theories with a light $Z'$ that mixes with the SM gauge bosons \cite{Frandsen:2011cg}.

\begin{figure}[!t]
\begin{center}
\includegraphics[width=0.45 \textwidth]{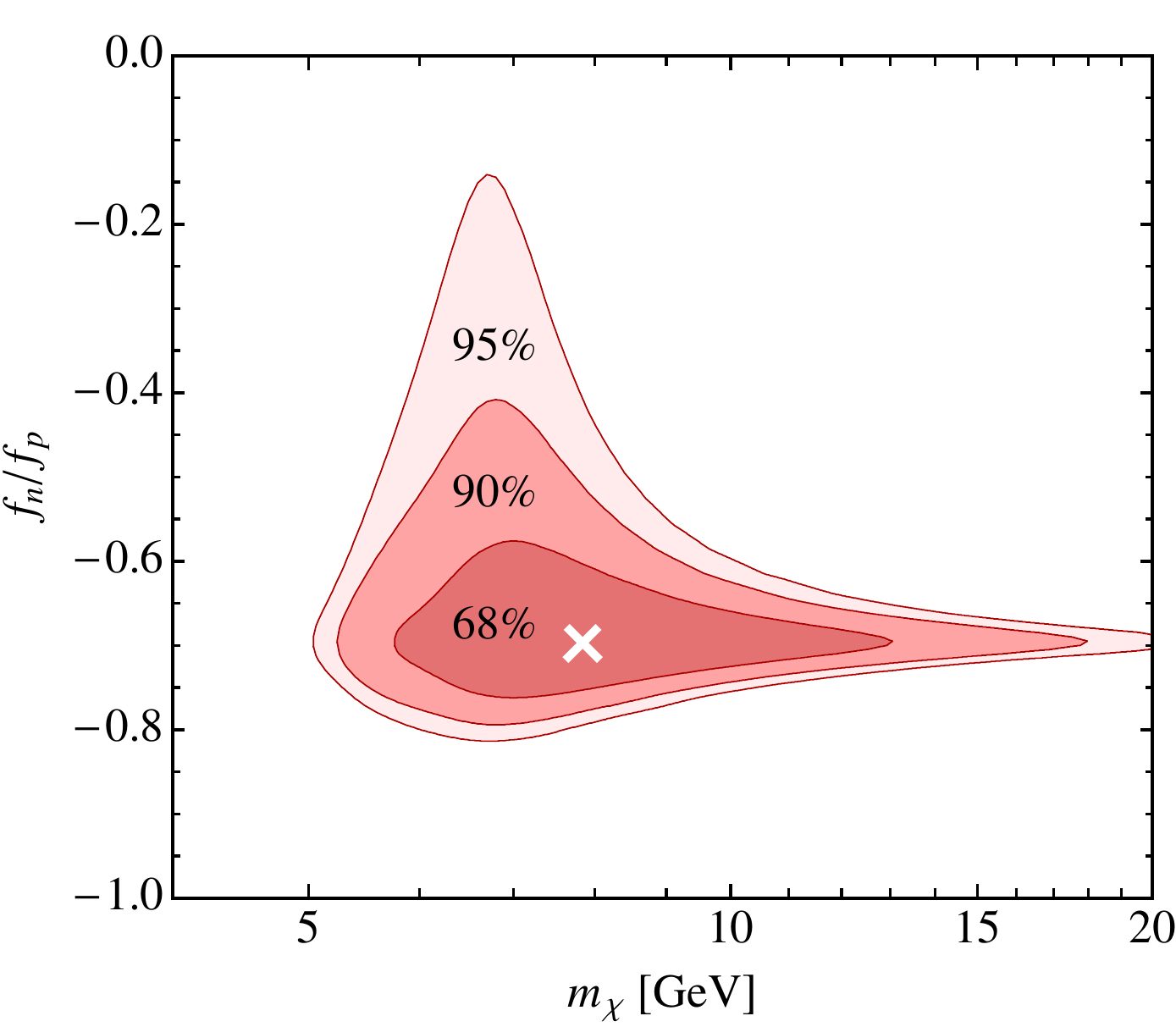}
\quad
\includegraphics[width=0.45 \textwidth]{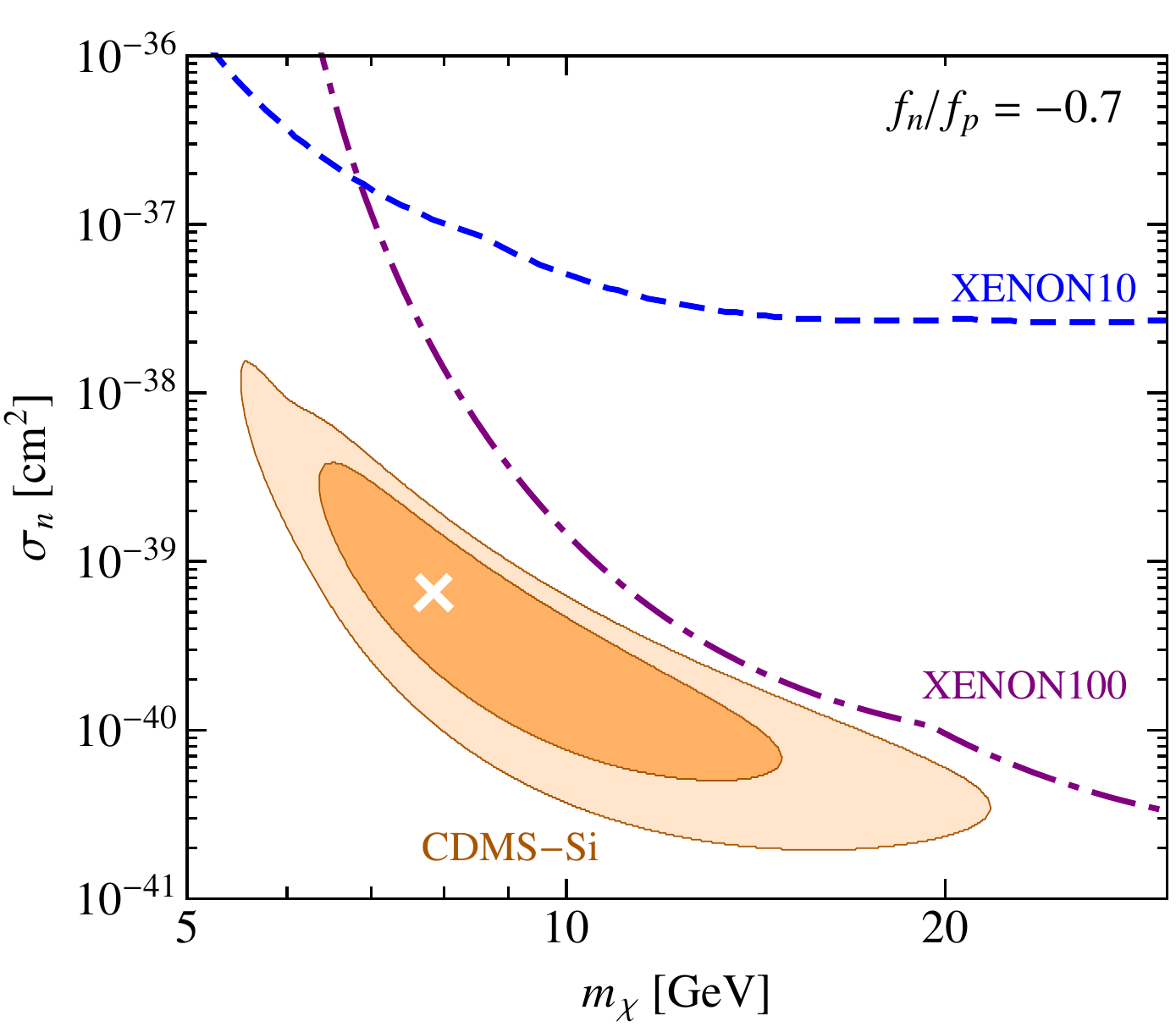}
\vspace{1mm}
\caption{\label{fig:fnfp}Isospin-dependent couplings. Left: Combined parameter estimation of $f_n / f_p$, $m_\chi$ and $\sigma_n$ (not shown) using a global maximum likelihood method (see text for details). As expected, there is a preference for $f_n / f_p = -0.7$ but the $2 \sigma$ confidence region extends up to $f_n / f_p \simeq -0.2$. Right: CDMS-Si allowed parameter region and XENON10/100 bounds for $f_n / f_p = -0.7$. In both plots, the best-fit point is indicated with a white cross.}
\end{center}
\end{figure}

To study this possibility, we scan simultaneously over $f_n / f_p$, $\sigma_n$ and $m_\chi$ and calculate the likelihood for each set of parameters, using the maximum likelihood method described in~\cite{Farina:2011pw}. In particular, we assume that the likelihood for XENON10 and XENON100 for a given set of parameters are given by $\mathcal{L}_\text{Xe} = \exp(-N_\text{max})$, where $N_\text{max}$ is the number of events expected in the largest interval determined using the maximum gap method. Since we use an extended maximum likelihood method to fit CDMS-Si, the minimum value of the likelihood depends on an arbitrary normalisation constant and does not carry physical significance. We  cannot therefore perform a goodness-of-fit analysis for our model, but we can infer confidence regions for parameter estimation. 

The results of this analysis are shown in the left panel of Fig.~\ref{fig:fnfp}. As expected \cite{Frandsen:2011cg}, the best-fit point corresponds to $f_n / f_p \simeq -0.7$, which strongly suppresses the bounds from XENON10/100, as shown in the right panel of Fig.~\ref{fig:fnfp}. Note that in this particular case, the strongest constraints on CDMS-Si arise from SIMPLE~\cite{Felizardo:2011uw} and the CRESST-II commissioning run~\cite{Brown:2011dp} (not shown). For  $f_n / f_p = -0.7$ these experiments require $\sigma_n \lesssim 10^{-39}\,\text{cm}^2$ at $m_\chi \simeq 10\,\text{GeV}$~\cite{Frandsen:2011gi} and therefore do not significantly constrain the CDMS-Si preferred region.

\begin{figure}[t]
\begin{center}
\includegraphics[width=0.45 \textwidth]{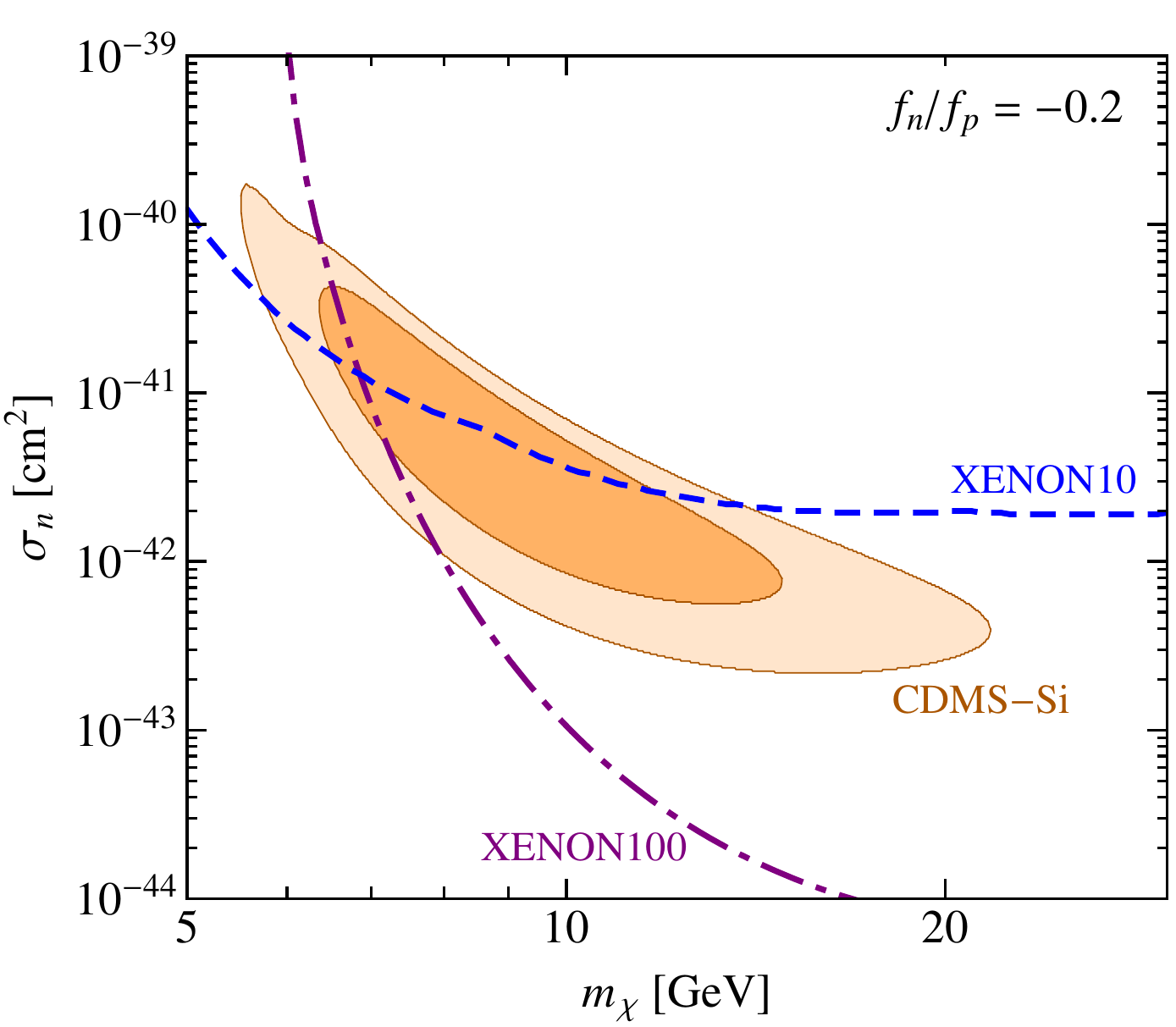}
\quad
\includegraphics[width=0.45 \textwidth]{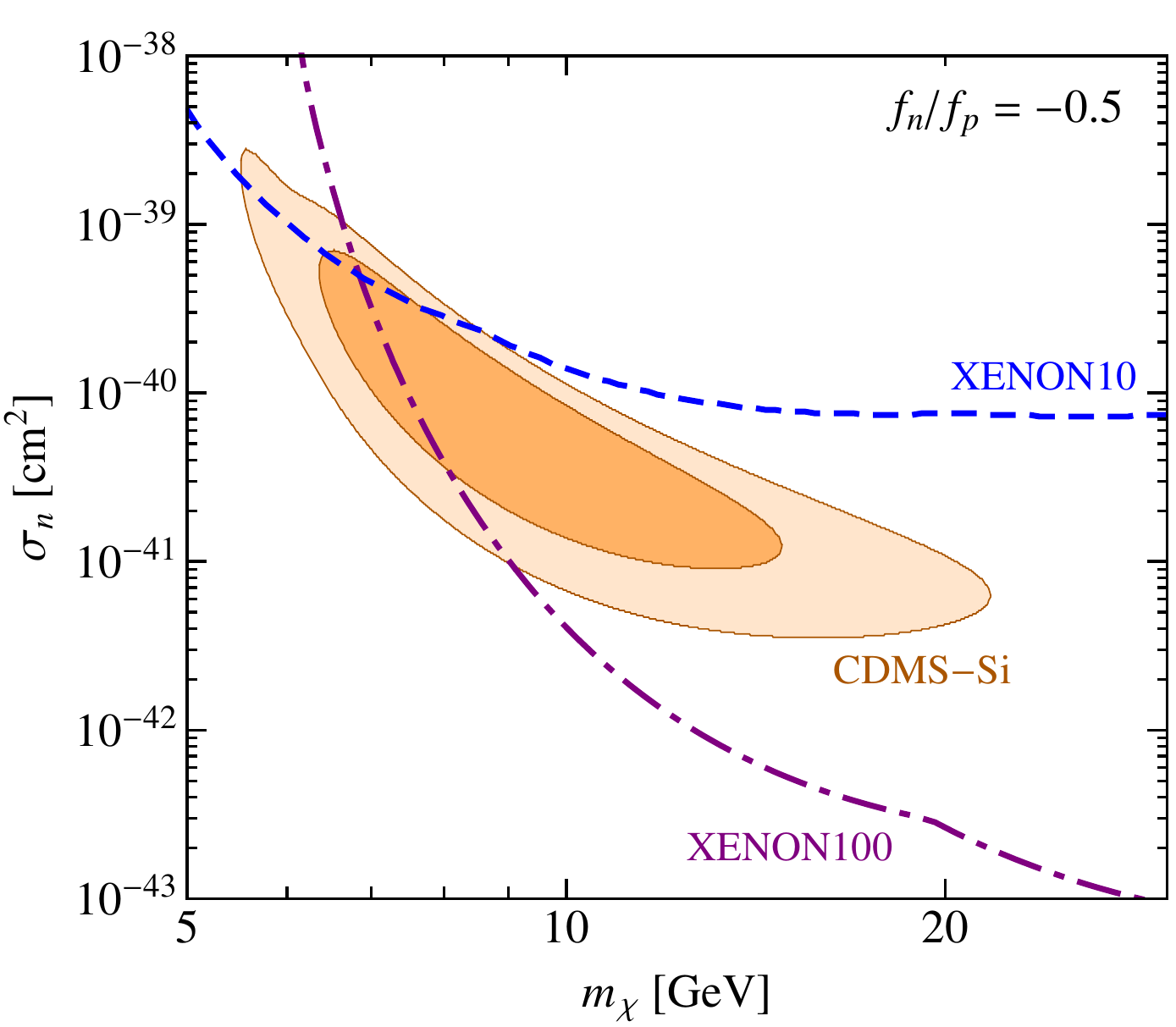}
\vspace{1mm}
\caption{\label{fig:isospin}Alternative choices for isospin-dependent couplings. No significant fine-tuning of $f_n / f_p$ is required to weaken the XENON10/100 bounds relative to CDMS-Si. Note the change of scales in these figures.}
\end{center}
\end{figure}

In spite of the preference for $f_n / f_p \simeq -0.7$, we observe that much larger values of $f_n / f_p$ still give a good fit to the data. At $1\sigma$ confidence level, we find $-0.76 < f_n / f_p < -0.58$ and the $2 \sigma$ confidence region extends up to $f_n / f_p \simeq -0.2$. To illustrate this point, we show the cases $f_n / f_p = -0.5$ and $f_n / f_p = -0.2$ in Fig.~\ref{fig:isospin}. We conclude that little fine-tuning is required to suppress the bounds from XENON10/100, in particular we do not require a precise cancellation of proton and neutron contributions and are therefore not sensitive to potential differences in the form factors as discussed in~\cite{Cirigliano:2012pq}. We would also emphasise that even though $f_n = f_p$ is strongly disfavoured compared to $f_n / f_p = -0.7$ when performing a parameter estimation, this does not imply that such a value cannot give a good fit to the data. In fact, as we have shown above, CDMS-Si and XENON10/100 \emph{are} compatible at the $90\%$ CL even for $f_n = f_p$. 

\section{Conclusions}\label{conclusions}

The report by the CDMS-II collaboration~\cite{Agnese:2013rvf} of 3 events consistent with nuclear recoils  from scattering of Galactic DM particles with a mass of $\sim 8.6$~GeV is both exciting and intriguing. This low-mass region has been attracting increasing theoretical interest as DM particles with such a mass can have a natural connection to baryons by virtue of sharing a primordial asymmetry. However there appears, \emph{prima facie}, to be a discordance with the upper limits placed by the XENON10/100 experiments.

We have demonstrated that in fact the XENON10/100 experiments do \emph{not} exclude the entire CDMS parameter region; in particular, the bound from XENON10 is significantly weaker than reported by the collaboration~\cite{Angle:2011th}. Moreover, this bound is rather sensitive to the (unknown) behaviour of the ionisation yield at the relevant low recoil energies.

Nevertheless there is tension between the CDMS results and the XENON10/100 bounds and we have shown explicitly that this is \emph{independent} of astrophysical uncertainties concerning the velocity distribution of halo DM e.g.\ the presence of a `debris flow'. Allowing a possible momentum or velocity dependence in the DM scattering cross-section will also not alleviate the tension as we have illustrated for both long-range and anapole interactions.

However the tension is reduced if scattering of DM on heavy targets like Xe is suppressed compared to light ones like Si, e.g.\ because DM couplings are isospin-dependent, leading to interference between scattering rates on protons and neutrons. As another example for enhancing scattering off light targets, we have briefly discussed exothermic DM.

In conclusion, the DM interpretation of the 3 events reported by the CDMS-II collaboration is not entirely excluded by XENON10/100. This mass region is very interesting theoretically and can be accessed with relatively small experiments employing light target nuclei. In view of our ignorance concerning the possible interactions of DM, it is essential that experimental searches employ as many different target nuclei as is possible. 

\subsection*{Note added in Proof}

The XENON10 Collaboration has issued an Erratum~\cite{Erratum} acknowledging the error in their published analysis~\cite{Angle:2011th} which we have pointed out in the present paper (\S~\ref{directdetection}). The corrected limit is stated to be in good agreement with our result shown in Fig.~\ref{fig:standard}.

\section*{Acknowledgements}

We thank C\'eline B\oe hm, Matthew Dolan, Joachim Kopp, Thomas Schwetz and Martin W.~Winkler for discussions and Milan Kundera for inspiration. MTF acknowledges a `Sapere Aude' Grant no.\ 11-120829 from the Danish Council for Independent Research. FK is supported by the Studienstiftung des Deutschen Volkes, STFC UK, and a Leathersellers' Company Scholarship at St Catherine's College, Oxford. SS acknowledges support by the EU Marie Curie Initial Training Network `UNILHC' (PITN-GA-2009-237920). KSH acknowledges partial support by the German Science Foundation (DFG) under the Collaborative Research Center (SFB) 676 `Particles, Strings and the Early Universe'.

\providecommand{\href}[2]{#2}\begingroup\raggedright\endgroup

\end{document}